\documentclass[a4paper,twoside,14pt]{article}

\usepackage[utf8]{inputenc}

\usepackage[english]{babel}
\usepackage{subeqnarray}
\usepackage{url}

\usepackage{graphicx,psfrag}
\usepackage{fixltx2e,amsmath}
\usepackage{xcolor}
\usepackage{amssymb}
\usepackage[final]{pdfpages}
\usepackage{bbold}
\usepackage{dsfont}
\usepackage{pgfplots}
\usepackage{subfig}
\usepackage{algorithm}
\usepackage{algorithmic,placeins}
\usepackage[font=small,labelfont=bf]{caption}

\usepackage[font={footnotesize}]{caption}

\usepackage[utf8]{inputenc}
\usepackage[T1]{fontenc}
\usepackage{xcolor,amsmath,amssymb}
\usepackage{geometry}
\usepackage{graphicx}
\usepackage[sort&compress,numbers]{natbib}
\usepackage{enumerate}
\usepackage{wasysym}
\usepackage[textsize=footnotesize]{todonotes}

\usepackage{hyperref}
\hypersetup{
    colorlinks=true,
    linkcolor=red,
    filecolor=magenta,      
    urlcolor=cyan,
}

\usepackage{pgfplots}

\definecolor{darkblue}{rgb}{0,0,0.6}
\definecolor{darkred}{rgb}{0.6,0,0}

\usepackage{hyperref}

\def\to{\rightarrow}

\newcommand{\beq}{\begin{equation}} \newcommand{\eeq}{\end{equation}}

\newcommand\be{\begin{equation}}
\newcommand\bea{\begin{eqnarray} \nonumber }
\newcommand\ee{\end{equation}}
\newcommand\eea{\end{eqnarray}}

\usepackage{authblk}

\usepackage[bitstream-charter]{mathdesign}

\usetikzlibrary{pgfplots.groupplots}

\graphicspath{{figures/}} 

\begin{document}

\title{Amorphous Order \& Non-linear Susceptibilities\\ in Glassy Materials}
\author[1]{Giulio Biroli}
\author[2]{Jean-Philippe Bouchaud}
\author[3]{Francois Ladieu}
\affil[1]{Laboratoire de Physique de l'Ecole Normale Sup\'erieure,
Universit\'e PSL, CNRS, Sorbonne
		Universit\'e, Universit\'e de Paris, F-75005 Paris, France}
		\affil[2]{CFM, 23 rue de l'Universit\'e, F-75007 Paris, France,
		\& Acad\'emie des Sciences, Quai de Conti, F-75006 Paris, France}
		\affil[3]{SPEC, CEA, CNRS, Universit\'e Paris-Saclay, CEA Saclay Bat 772, F-91191 Gif-sur-Yvette Cedex, France.}
\date{December 2020}
\maketitle

\abstract{We review 15 years of theoretical and experimental work on the non-linear response of glassy systems. We argue that an anomalous growth of the peak value of non-linear susceptibilities is a signature of growing ``amorphous order'' in the system, with spin-glasses as a case in point. Experimental results on supercooled liquids are fully compatible with the RFOT prediction of compact ``glassites'' of increasing volume as temperature is decreased, or as the system ages. We clarify why such a behaviour is hard to explain within purely kinetic theories of glass formation, despite recent claims to the contrary.} 

\section{Introduction: RFOT \& Amorphous Order}

In the bestiary of material science, glasses play the role of hybrid creatures: half-liquid, half-solid, like the centaur or the mermaid. A glass has the structure factor of a liquid, yet it does not flow and responds elastically to shear deformations -- at least on time scales much shorter than the relaxation time $\tau$ of the system, which itself increases extraordinarily fast as temperature is decreased. As an illustration, the relaxation time of ``fragile glasses'' (such as Ortho-Terphenyl) increases by a factor $10^{10}$ as temperature drops by a mere $10\%$. 

From a general point of view, a non-zero static shear modulus is necessarily associated with a loss of ergodicity, and thus a transition into a state where the dynamics is no longer able to probe the entire phase-space. The fundamental question that has riveted theoreticians for decades is whether the physics of glasses is indeed driven by an underlying phase transition into an ergodicity-broken state characterized by some amorphous long-range order, or whether the dramatic slowdown is purely of kinetic origin, with no particular thermodynamic signature. 

The idea of amorphous order sounds like an oxymoron, but in fact spin-glasses \cite{binderyoung} offer a blueprint for such a scenario. 
Randomly interacting spins (some pairs favouring alignment while others favour anti-alignment) are known to undergo a phase transition towards a frozen, ``spin-glass'' state below some critical temperature $T_c$. Each spin then points in a random direction, but this direction remains constant in time -- whereas at higher temperature $T>T_c$, it flips randomly across time. Much as in glasses, instantaneous snapshots of the spin configurations seem featureless both above and below $T_c$. But whereas there is no long range transmission of information above $T_c$, the spin-glass phase is {\it rigid}, in the sense that localized perturbations have a long range effect on the system -- much like the free-energy per particle of rigid bodies depends on the shape of its boundaries.
In this sense, the spin-glass is indeed characterized by long-range amorphous order.  

Said differently, whereas spins respond more or less independently in the high temperature phase, they respond {\it collectively} in the spin-glass phase. As the transition temperature is approached from above, the collective nature of the response increases, and in fact diverges at $T_c$. But contrarily to ferromagnets where all spins point in the same direction, the linear magnetic susceptibility of a spin-glass does {\it not} diverge at $T_c$, because the correlated clusters (whose spins are temporarily inter-locked by interactions) carry a total magnetisation proportional to the square-root of their volume, and not to their volume (see Eq. \eqref{eq2} below for a more precise statement). However, remarkably, all higher order {\it static} non-linear susceptibilities diverge at $T_c$, and directly elicit the growth of amorphous order as the spin-glass phase is approached \cite{binderyoung,Lev86,janus}. 

How much of this phenomenology is shared by supercooled liquids and other glassy materials? If glasses are simply hyper-slow liquids, as Kinetically Constrained Models (KCM) posit \cite{Ritort_Sollich,KCM}, one should not expect any anomalous response to an external field. But if the glass transition is intimately related to the growth of amorphous order -- now in the sense of molecules inter-locked in particular relative positions and orientations -- then we should expect, as we have indeed observed experimentally, a strong increase of non-linear susceptibilities as the system slows down. 

From a theoretical point of view, the deep analogy between glasses and spin-glasses\footnote{initially anticipated by Phil Anderson, who wrote that {\it some -– but not all –- transitions to rigid, glass-like states, may entail a hidden, microscopic order parameter which is not a microscopic variable in any usual sense, and describes the rigidity of the system.}} finds its roots in the landmark series of papers by Ted Kirkpatrick, Dave Thirumalai and Peter Wolynes in the mid 80's \cite{KTW1,KTW2,KTW3}. Based on the solution of a family of mean-field models of spin-glasses, they proposed their ``Random First Order Transition'' (RFOT) theory \cite{KTW3,RFOT1,RFOT2}, which appears to capture all the known phenomenology of supercooled liquids, in particular:
\begin{itemize}
    \item The existence of a cross-over temperature $T^\star$ below which the relaxation function exhibits a plateau as a function of time; such plateau is associated with the appearance of local rigidity (also often called ``cage formation'') with a non-zero high frequency shear modulus $G_{\text{hf}}$;
    \item The existence of an ideal glass transition (Kauzmann) temperature $T_K$ where the configurational entropy vanishes;
    \item An Adam-Gibbs-like correlation between the logarithm of the relaxation time and the inverse of the configurational entropy \cite{Ada65}. 
\end{itemize} 

More precisely, the RFOT theory envisages the glass state as a mosaic of ``glassites'' (i.e. locally frozen clusters), with a size $\ell$ inversely related to the configurational entropy -- and thus diverging as $T \downarrow T_K$ \cite{KTW3,Bou04}. Being of finite size, the life-time $\tau$ of these glassites is also finite, but grows exponentially with $\ell$. These glassites are rigid, in the sense that boundary conditions are able to lock all inside molecules around a fixed position \cite{Bou04,Cavagna}. Hence, glassites respond elastically to an external shear for times less than $\tau$, with a shear modulus $G_{\text{hf}}$, but start flowing for times larger than $\tau$ when the local order finally unravels. Using a simple Maxwell model, the viscosity $\eta$ of the supercooled liquid is thus $\eta \approx G_{\text{hf}} \tau$. 

Note that the relative positions and orientations of the molecules inside  RFOT glassites are frozen (apart from small and fast ``cage'' oscillations), not because of kinetic constraints but because of actual inter-molecular forces that favour specific low energy configurations and drive the system towards a thermodynamic glass when $\ell \to \infty$. These forces will, by the same token,  correlate the re-orientations of the dipoles induced by an oscillating electric field within each glassite, provided the frequency $\omega$ of the electric field is somewhat larger than $\tau^{-1}$. Hence, we expect that glassites respond {\it collectively} to an external field, amplifying, as for spin-glasses, all non-linear susceptibilities. At variance with spin-glasses, however, this collective response only holds when $\omega \tau \gtrsim 1$. At lower frequencies, glassites melt and all collective effects are lost \cite{Bou05}. 

This argument suggests that, if the RFOT picture is correct, all non-linear susceptibilities should peak for $\omega \tau \sim 1$, with a peak amplitude growing with the moment of the frozen dipole, and thus with the size of the glassite itself. On the other hand, purely kinetic effects cannot lead by themselves to growing non-linear susceptibilities. As we shall show later, in this case one should expect a peak of non-linear susceptibilities that shift to lower frequencies as $\tau$ increases, but with a roughly constant amplitude, independent of temperature.  

In the following sections, we formulate more precise theoretical statements about the behaviour of non-linear susceptibilities in correlated systems, and review recent experimental results that appear to confirm the analogy between glasses and spin-glasses and the presence of amorphous order that extends to larger and larger length scales as the system slows down and/or ages. We discuss more formally why Kinetic Constraints cannot account for the experimental results -- although this does not mean that the basic tenet of KCMs, i.e. dynamic facilitation, is irrelevant \cite{Xia,CW,crumbling}. Finally, we discuss open issues and directions for future research, both theoretical and experimental.   

\section{Non-Linear Susceptibilities \& Amorphous Order: Theoretical Results}

We first present a phenomenological argument that allows one to understand the growth of susceptibilities in correlated systems. This argument can in fact be fully justified in the proximity of a phase transition, where the correlation length becomes very large \cite{Alb16}. Suppose that $N_{\text{corr}} = (\ell/a)^{d_f}$ molecules are strongly correlated over a distance $\ell$, with $a$ the molecular size and $d_f$ the fractal dimension of the correlated clusters. The corresponding dipoles are then essentially locked together during a time $\tau$, resulting in a total polarisation of the cluster scaling as $N_{\text{corr}}^\zeta$, with $\zeta = 1$ for ferromagnetic interactions, and $\zeta=1/2$ for completely random interactions, as in spin-glasses.\footnote{Intermediate values of $\zeta$ could be expected near a tri-critical point separating a ferromagnetic, paramagnetic and spin-glass phases.} One expects that in the presence of an external field $E$ oscillating at frequency $\omega \sim \tau^{-1}$, the dipolar degrees of freedom of the molecules contribute to the polarisation per molecule $p$ as:
\begin{equation} \label{scaling}
p = \mu \frac{{(\ell /a)^{\zeta d_f}}}{(\ell /a)^d} \mathcal{F}\left(\frac{\mu E {(\ell /a)^{\zeta d_f}}}{kT}\right),
\end{equation}
where $\mu$ is an elementary dipole moment, $\mathcal{F}$ a scaling function such that $\mathcal{F}(-x)=-\mathcal{F}(x)$, and $d=3$ the dimension of space.\footnote{In the spin-glass case, Eq. \ref{scaling} is in fact equivalent to the scaling arguments of Ref. \cite{Lev88}.}

Expanding in powers of $E$, one readily finds that the field induced polarisation $p$ is given by:
\begin{eqnarray} \label{eq2}
\frac{p}{\mu} =  \mathcal{F}'(0) \left(\frac{\ell}{a} \right)^{2 \zeta d_f - d} \left(\frac{\mu E}{kT}\right) +
\frac{1}{3!} \mathcal{F}^{(3)}(0) \left(\frac{\ell}{a} \right)^{4 \zeta d_f - d} \left(\frac{\mu E}{kT}\right)^3 +  \frac{1}{5!}\mathcal{F}^{(5)}(0) \left(\frac{\ell}{a} \right)^{6 \zeta d_f - d} \left(\frac{\mu E}{kT}\right)^5 + \dots
\end{eqnarray}
\begin{itemize} 
\item When $2 \zeta d_f > d$, which is the case of usual ferromagnets for which $\zeta=1$,\footnote{In this case, $2d_f=d+2 - \eta$, where $\eta$ is the standard critical exponent for the decay of correlations \cite{Coniglio}. The resulting divergence of the linear susceptibility $\chi_1$ as $\ell^{2-\eta}$ can also be obtained directly using the Fluctuation-Dissipation Theorem.} one finds that the linear susceptibility $\chi_1$ diverges when $\ell$ increases, revealing an incipient ferromagnetic order. 
\item When $\zeta=1/2$, on the other hand, the linear susceptibility cannot diverge since $d_f \leq d$. Therefore, amorphous order cannot be revealed by the linear susceptibility. But the second term,
contributing to the third-order susceptibility, does grow with $\ell$ provided $2 d_f > d$ (or more generally when $4 \zeta d_f > d$). Such a divergence is indeed observed as the spin-glass transition is approached, see Fig. 1 for experimental results and \cite{janus} for state of the art numerical results. This is a clear indication that the weird concept of amorphous order is indeed relevant for this class of materials: the spin-glass phase is a quite an exotic phase of matter.  
\end{itemize}

\begin{figure}[t]
\centering
\includegraphics[width = 8cm]{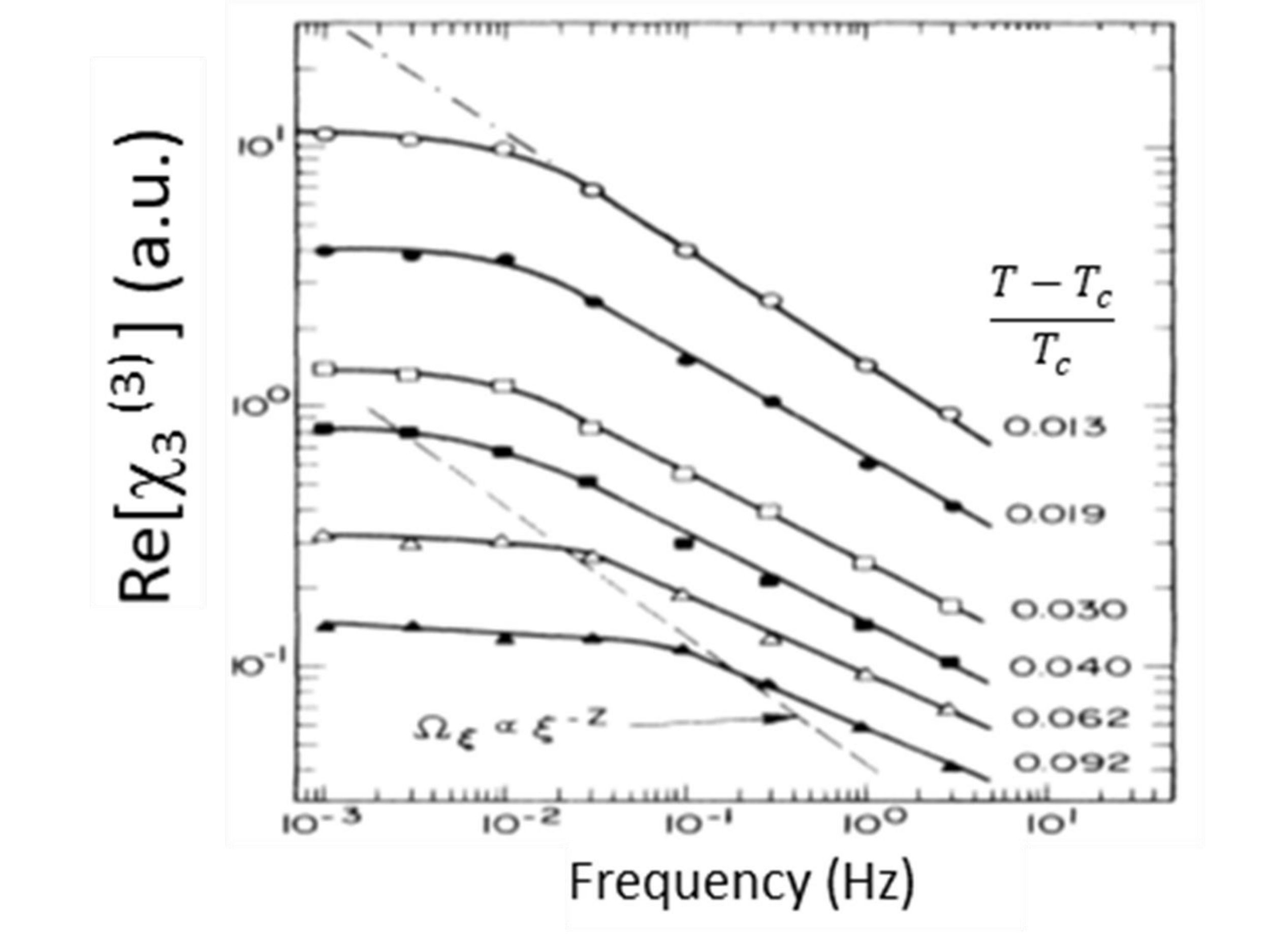}
\caption{Adapted from Ref. \cite{Lev86}. Third-harmonic susceptibilities in the Ag:Mn spin glass \cite{Lev86}: the static value of $\chi_3^{(3)}$ diverges when approaching  the critical temperature $T_c \simeq 2.94$~K
 \cite{Lev86}. At a given $T > T_c$, $\chi_3$ as a plateau below the characteristic frequency $1/\tau(T)$ with a power law decrease at high frequencies.}
\label{fig1}
\end{figure}

Whereas $d_f < d$ close to the spin-glass transition,\footnote{For 3D Ising spin-glasses, experiments indicate that $d_f \approx 2.4$ \cite{Lev88} whereas large scale numerical simulations give $d_f \approx 2.7$ \cite{janus}.} RFOT theory suggests that for supercooled liquids, glassites are compact ($d_f=d$) \cite{KTW3,WolynesSchmalian}. Assuming further that dipoles are randomly oriented in a typical frozen state leads to $\zeta=1/2$ as in spin glasses, one gets a third-order peak susceptibility scaling as
\begin{equation}\label{chi3_peak}
    |\chi_{3,\text{peak}}| \propto \ell^{4 \zeta d_f - d} = \ell^{d} = N_{\text{corr}} a^d \qquad \text{when} \qquad \omega \tau \sim 1
\end{equation}
 The third term of Eq. \ref{eq2} reveals that the fifth-order susceptibility $\chi_{5,\text{peak}}$ should diverge as $\ell^{6 \zeta d_f - d}$. Hence, the joint measurement of $\chi_3$ and $\chi_5$ provides -- in principle -- a direct way to estimate $\zeta d_f$ experimentally, through the following relation:
\begin{equation} \label{def-gamma}
|\chi_{5,\text{peak}}| \propto |\chi_{3,\text{peak}}|^{\gamma}; \qquad \gamma=\frac{6 \zeta d_f-d}{4 \zeta d_f-d}. 
\end{equation}
where $\gamma$ is equal to $2$ when the dynamically correlated regions are compact ($d_f=d$) and frozen dipoles are randomly oriented ($\zeta=1/2$).

Let us now briefly summarize the experimental results on the non-linear susceptibility of super-cooled liquids obtained in the last decade. 

\section{Experimental Results} 

The measurement of non-linear dielectric susceptibilities of supercooled liquids such as glycerol or propylene carbonate is not easy.\footnote{For an early attempt, see L. Wu, Phys. Rev. B 43, 9906 (1991).} Yet it has now been reliably achieved by several groups \cite{Ric06,Cra10,Bru11,Bru12,Bau13a,Bau13b,Cas15,Ric17,Lun17,Gad17}. Measuring $\chi_5$ is even more difficult, but after many years of  efforts two groups (in Saclay and in Augsburg) have reported similar results that they published in a joint paper \cite{Alb16}. Here, we summarize the salient features of these results, without doing justice to the intricate details of the experiments. 

In order to elicit the temperature dependence of collective effects, one introduces the following dimensionless quantities:
\begin{equation}
X_3 := \frac{k T}{\epsilon_0 \Delta \chi^2 a^3} \chi_3, \qquad
X_5 := \frac{(k T)^2 }{\epsilon_0^2 \Delta \chi^3 a^6} \chi_5
\label{eqX5}
\end{equation}
where $\epsilon_0$ is the permittivity of free space, $\Delta \chi = \chi_1(\omega=0) -\chi_1(\omega \to \infty)$ is the ``dielectric strength'' (removing high frequency contributions irrelevant for the glass physics), $a^3$ is the molecular volume and $k$ is the Boltzmann constant. The main advantage of these dimensionless non-linear susceptibilities is that
in the trivial case of an ideal gas of dipoles, the values of $X_3$ and $X_5$ are \textit{independent} of temperature once plotted as a function of $\omega \tau(T)$. Hence, their experimental variation can be ascribed to the non-trivial dynamical correlations in the supercooled liquid \cite{Bou05,Lad12}. 

The experimental results on $X_3$ and $X_5$ for several supercooled liquids can be summarized as follows:
\begin{itemize}
    \item For a fixed temperature $T$ in the supercooled regime, $X_3$ has a humped shape as a function of frequency, as shown in Fig.  2,  with a maximum obtained at a frequency $\approx 0.2\, f_\alpha$ where the relaxation frequency is $f_\alpha = 1/(2\pi \tau(T))$. 
    \item The height of the peak of $X_3$ increases as the temperature decreases, whereas $X_3(\omega \to 0)$ is roughly constant, see Fig.  2. 
    \item The rescaled fifth-order susceptibility $X_5$ is also hump-shaped but with a much stronger increase in peak value (compared to $X_3$) as temperature decreases, as shown in Fig. 3. 
    \item The relation between $X_{3}^{\text{peak}}$ and $X_{5}^{\text{peak}}$ is compatible with Eq. (\ref{def-gamma}) with $\gamma \approx 2.0 \pm 0.5$, see Fig.  \ref{fig4}. 
    \item For frequencies larger than $f_\alpha$, $X_3(\omega)$ and $X_5(\omega)$ decrease as clean power-laws, symptomatic of a broad spectrum of relaxation times \cite{Bru12}.
\end{itemize}
\begin{figure}[!h]
\centering
\includegraphics[width = 10cm]{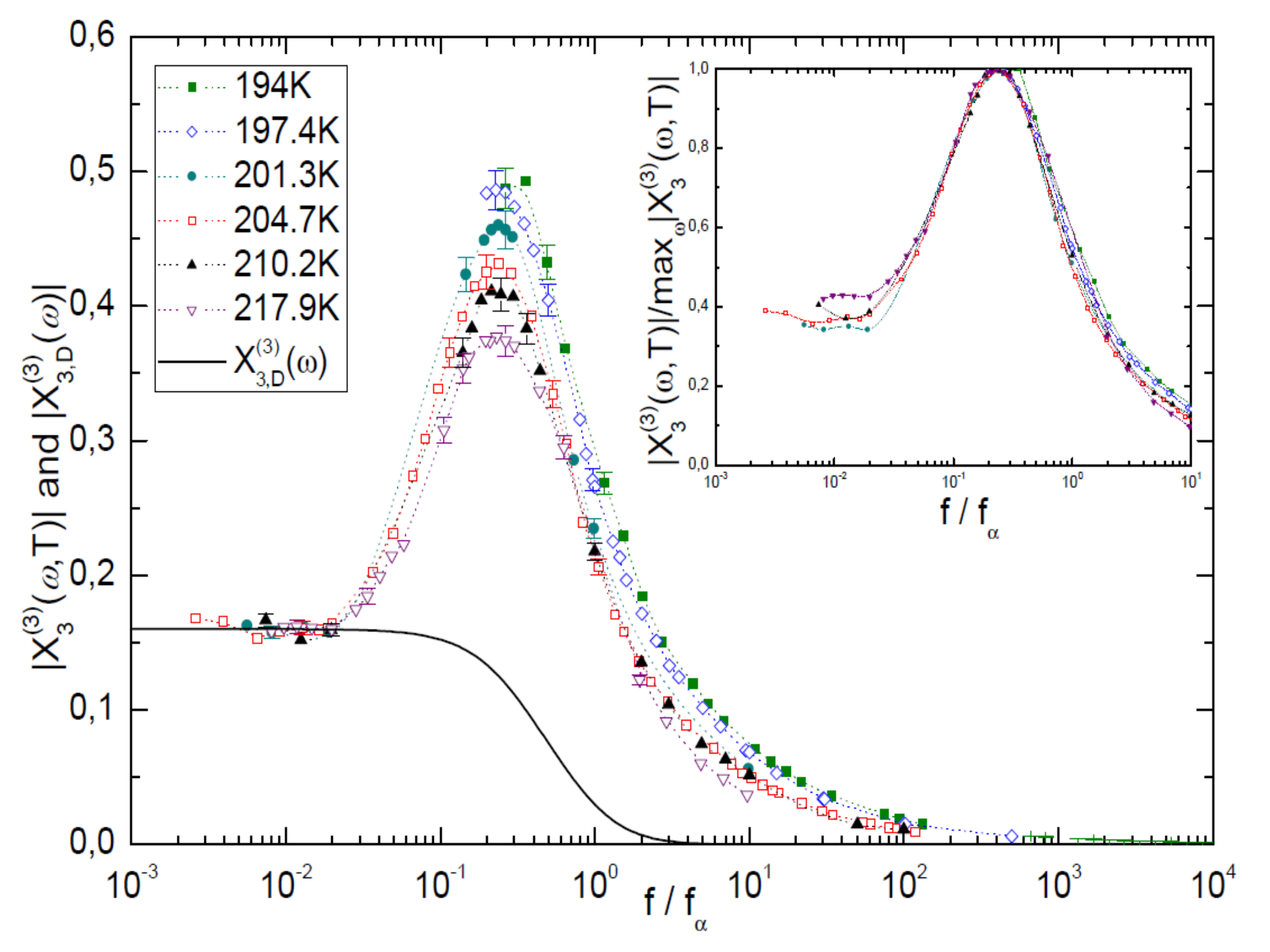}
\caption{Adapted from Ref. \cite{Bru11}. Dimensionless third-harmonics susceptibility $X_3^{(3)}$ in glycerol (with a glass transition $T_g \simeq 188$K) as a function of the reduced frequency $f/f_\alpha$. $X_3^{(3)}$ has a plateau at very low frequencies which does not appreciably depends on $T$. By contrast the maximum value of $X_3^{(3)}$ systematically increases upon cooling. The solid line $X_{3,D}^{(3)}$ is the dimensionless cubic susceptibility in a ideal gas of dipoles, and does not depend at all on T when plotted as a function of $f/f_\alpha$: this ``trivial'' case thus serves as a benchmark to elicit the glassy correlations (see \cite{Alb16} for more details). {\textit Inset:} same data excepted that $X_3^{(3)}$ is normalized by its maximum value over frequency.}
\label{fig2}
\end{figure}

\begin{figure}[!h]
\centering
\includegraphics[width = 10cm]{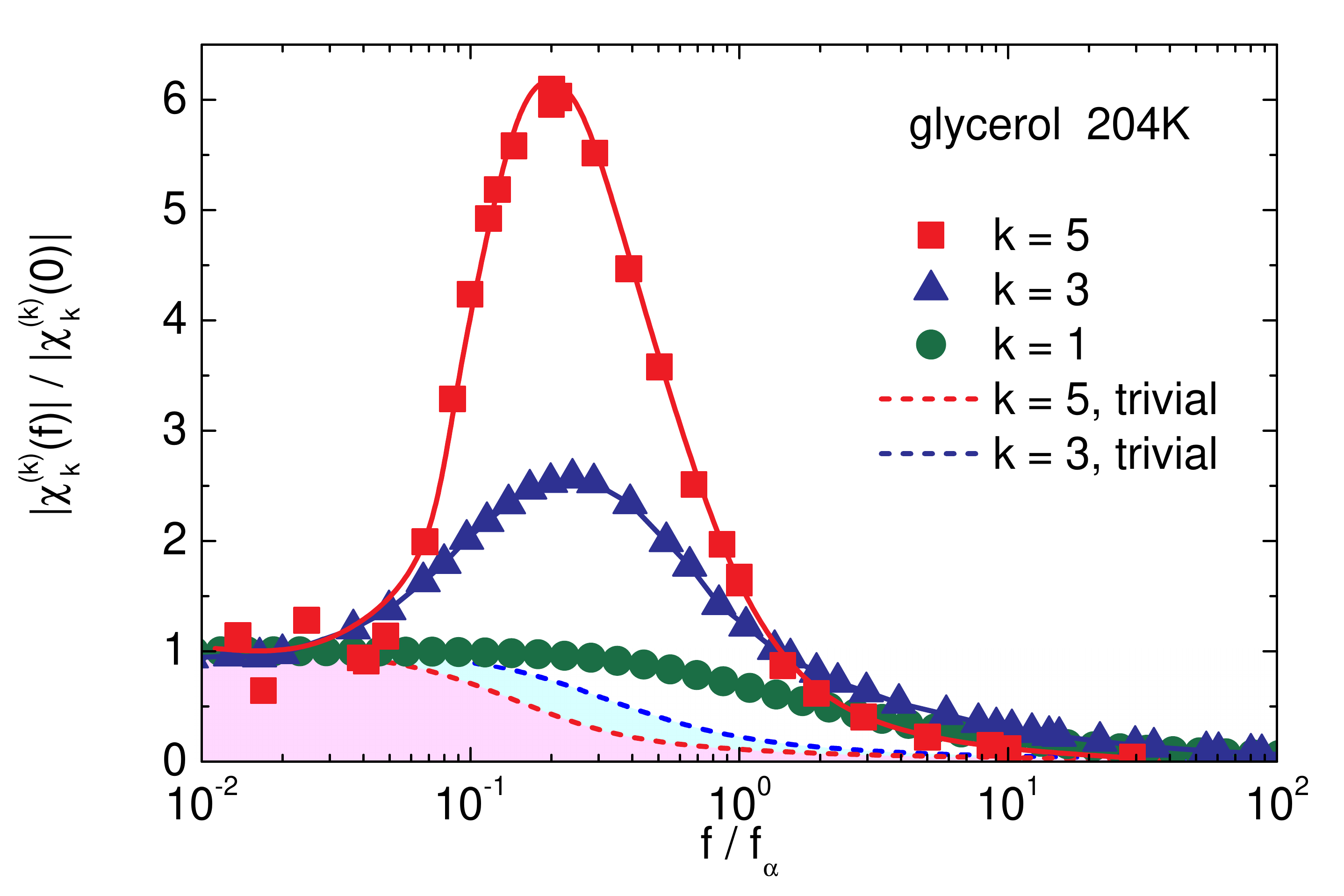}
\caption{Adapted from Ref. \cite{Alb16}. Comparison of the susceptibilities of various orders $\chi_k^{(k)}$ (scaled by their value at zero frequency) in glycerol at $T=204$K $\simeq T_g + 16$K. In practice the cubic (resp. fifth order) susceptibility has been measured by monitoring the response at $3 \omega$ (resp. $5 \omega$). For convenience the linear susceptibility has been denoted here $\chi_1^{(1)}$. Similar results were obtained in propylene carbonate \cite{Alb16}.  Two points are noteworthy: {\textit (i)} the humped shape in frequency in only present for nonlinear susceptibilities (in fact, their modulus); \textit{(ii)} the humped shape is much more pronounced for $\chi_5^{(5)}$ than for $\chi_3^{(3)}$. For comparison the dashed lines show the corresponding curves for the ``trivial'' case of an ideal gas of dipoles (see \cite{Alb16} for more details).}
\label{fig3}
\end{figure}

\begin{figure}[!h]
\centering
\includegraphics[width = 12cm]{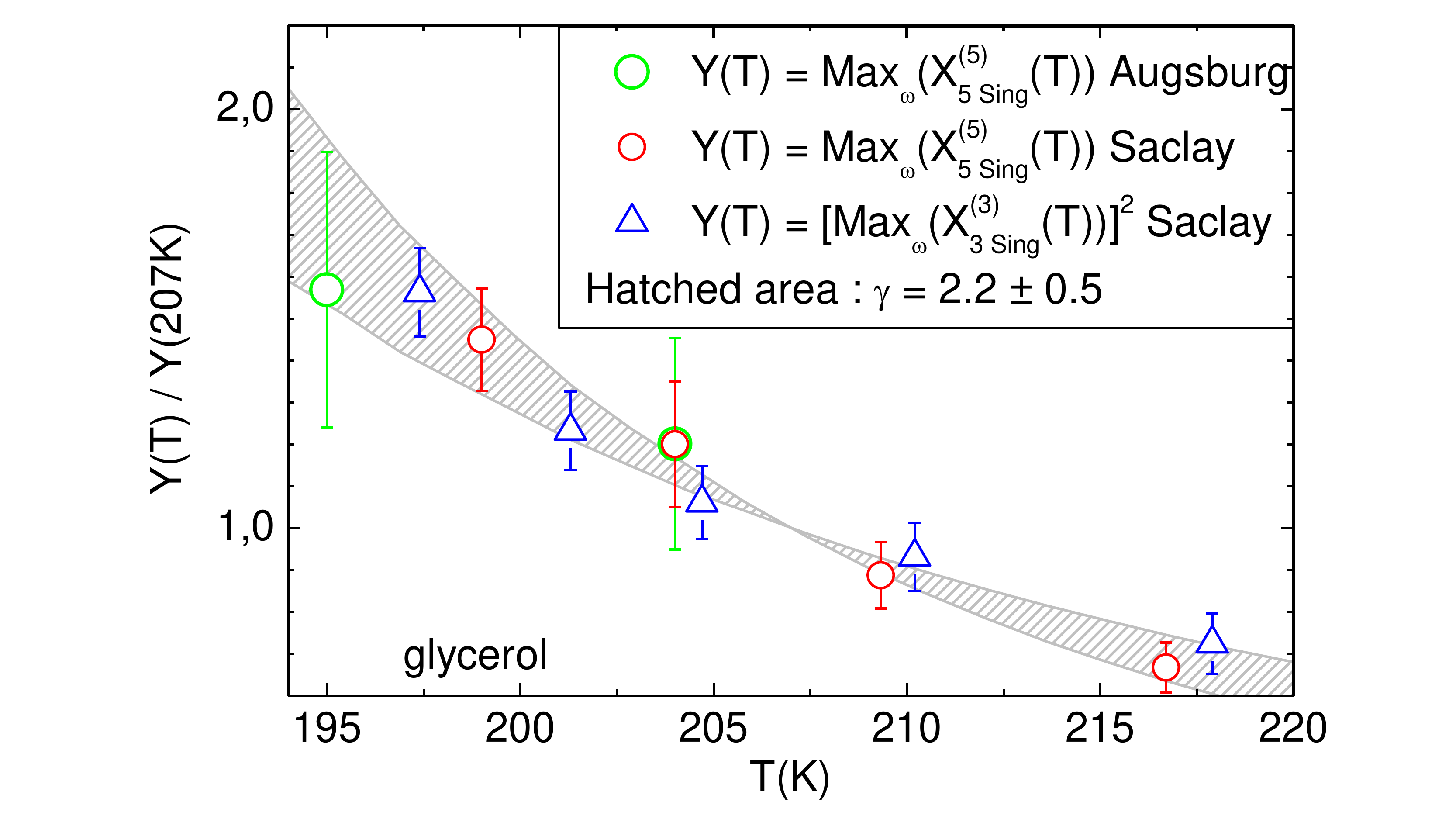}
\includegraphics[width = 12cm]{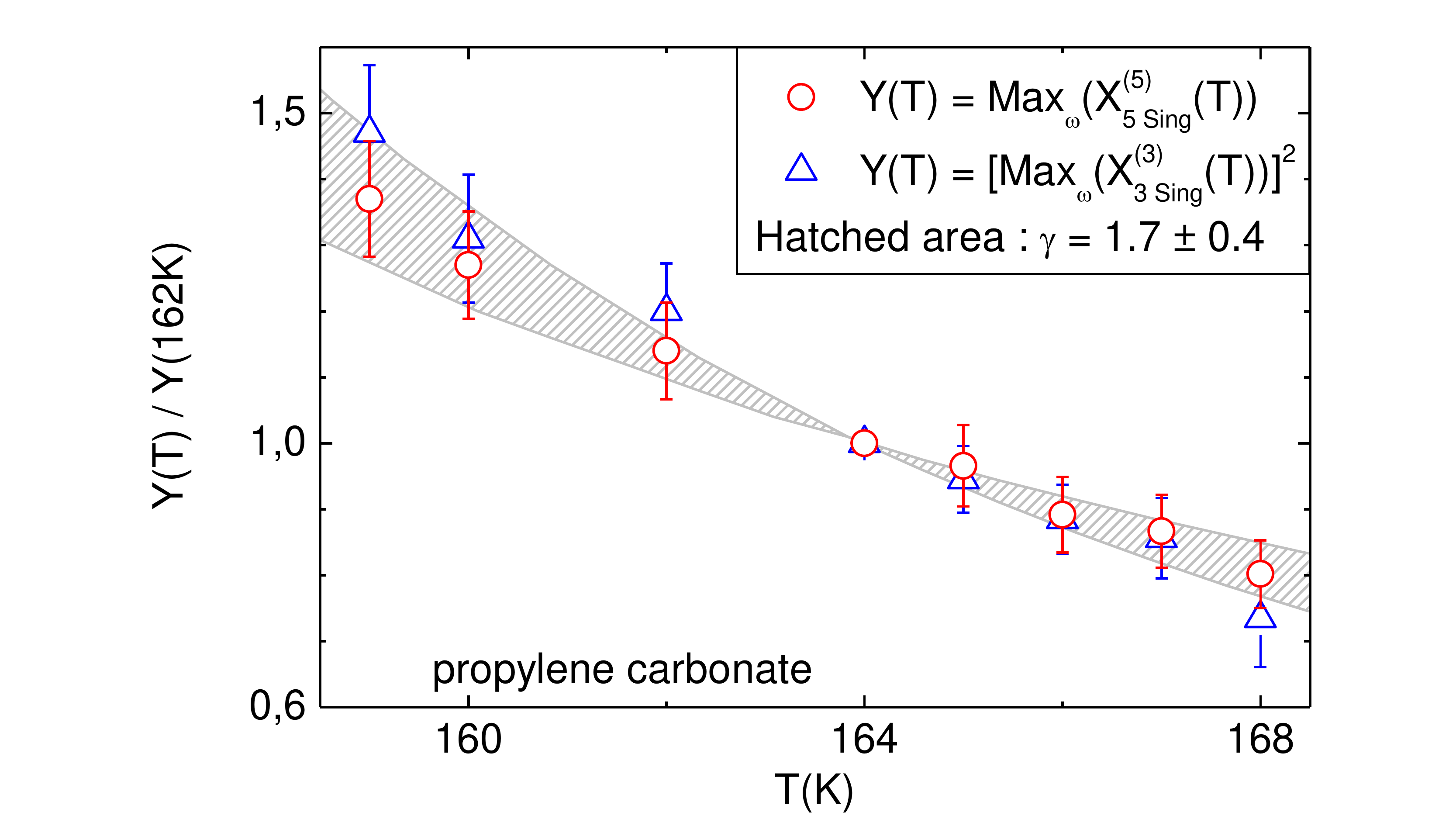}
\caption{Adapted from Ref. \cite{Alb16}. Temperature evolution of the singular parts of fifth and third order responses. All quantities are normalized at a given temperature, namely $207$~K for glycerol, upper panel, and $164$~K for propylene carbonate, bottom panel. This allows one to determine the exponent $\gamma$ relating $\vert X_5 \vert$ and $\vert X_3 \vert^\gamma$ and to conclude that the amorphously ordering domains are compact (see text). The hatched areas represent the uncertainty on $\gamma$. }
\label{fig4}
\end{figure}

Aging experiments were also performed, with the conclusion that the amplification of $\chi_3$ also increases with the age of the system, at fixed temperature \cite{Bru12}. This is in agreement with the idea that amorphous order, remarkably, propagates over larger and larger distances as the system ages.

All these features are compatible with the RFOT picture of temporarily frozen glassites. The fact that $X_3(\omega)$ is humped shape corresponds to the unraveling of the amorphous order at long times (i.e. $\omega \to 0$); the fact that $X_3(\omega \sim \tau^{-1})$ grows as temperature is decreased is compatible with Eq. \eqref{chi3_peak} with $\ell$ (or $N_{\text{corr}}$) increasing as the system is cooled down. Furthermore, the fact that $\gamma$ is close to $2$ means that (assuming $\zeta=1/2$) glassites are compact objects with $d_f=d=3$, as predicted by theory, at least  sufficiently deep below $T^\star$ \cite{WolynesSchmalian}. 
 
Finally, the power-law dependence of $X_3$ and $X_5$ for $\omega\tau \gg 1$ suggests the existence of a broad distribution of relaxation time for amorphous order, possibly related to a distribution of the size of glassites. A quantitative explanation of this power-law behaviour, and the value of the measured exponents, is however not available at this stage.\\  

Following up on these non-linear dielectric susceptibilities experiments, non-linear mechanical responses have been studied in glassy colloids \cite{fuchs}. In that case, one can probe experimentally a milder regime of super-cooling in which the growth of nonlinear responses can be well explained by Mode Coupling Theory and its extensions \cite{fuchs,Tar10}.

\newpage 

\section{\label{secKCM} Non-Linear Susceptibilities \& Kinetic Constraints}

Kinetically Constrained Models encode the view that the main physical ingredient explaining the spectacular slowing down of glasses is the rarefaction of point defects that act as facilitators for structural rearrangements \cite{KCM}. Relaxation is postulated to take place only when one defect passes close-by. In this scenario thermodynamics plays only a minor role: The system is simply a liquid that cannot flow, only because of kinetic constraints. There is no driving force towards any kind of locally preferred structure or amorphous order. In between two relaxation events, the system is in a state typical of the liquid phase, i.e. without any special type of correlations between molecules. 

The density $c(T)$ of facilitation defects is assumed to be small (for dynamics to be sluggish) and to go down quickly with temperature:
\begin{equation}
    c(T) = c_0 e^{-J/kT},
\end{equation}
where $J$ is the energy needed to create a defect. This translates, for sufficiently complex facilitation rules \cite{Ritort_Sollich,KCM}, into a relation between relaxation time $\tau$ and density of the form
\begin{equation}
    \tau(T) \sim [c(T)]^{-z(T)},
    \label{taukcm}
\end{equation}
with a temperature dependent exponent $z(T)$ that increases as temperature decreases, accounting for the super-Arrhenius dependence of $\tau(T)$ on temperature. Ref. \cite{KCM} argues that $z(T)=(T_o-T)/T$, where $T_o$ is an ``onset'' temperature where local rigidity appears (in fact quite analogous to the incipient rigidity temperature $T^\star$ mentioned above, which already suggests that rather non-trivial correlations must underpin the KCM picture, see below).
Furthermore, each independent defect 
unlocks the dynamics for a cluster of size $\sim c^{-1/d}$. This happens sequentially (when the defect passes by) for the molecules belonging to the same cluster. In consequence, KCMs predict the presence of dynamical heterogeneities, with molecules in clusters of size $\sim c^{-1/d}$ unlocking in a correlated way.\footnote{In some more refined theories, the unlocked clusters are fractal and $c^{-1/d}$ is replaced by $c^{-1/d_f}$ \cite{KCM,speck}.}

Note however that while molecules within each cluster do unlock in a correlated fashion (due to the presence of the same defect wandering about), these molecules are still basically free to reorganize each in their own way. So in the presence of an oscillating electric field, each dipole carried by these molecules reorients in a direction that is roughly {\it independent} of the direction in which other dipoles in the cluster decide to point. Contrarily to the case of spin-glasses, or of frozen RFOT glassites, there is no spatial propagation of information when slowing down is purely of kinetic origin. 
Another way to see that KCMs {\it do not} lead to super-dipoles responding to the oscillating field is to go back to Eq. (1) and ask what is the energy scale that couples to the field when thermodynamics is trivial and dynamics is unlocked by wandering defects. Since there is no spatial correlation and relaxation takes place sequentially, during an oscillation the external field sequentially couples to the local dipoles encountered by the defect. Hence, for KCM-like dynamics, the scaling function of Eq. (1) can only a function of $\mu E/kT$: there is no amplification of the effect of the field in this case, at variance with what happens in the presence of amorphous order extending over some length scale $\ell$.  

In conclusion, while one still expects all susceptibilities to peak around $\omega \sim \tau^{-1}$ in KCMs, there is no possibility for an increase of the peak value of $X_3$ or $X_5$ in the absence of any thermodynamic rigidity, in spite of claims to the contrary.  For example, T. Speck \cite{speck} has recently argued that the experimental result $X_5 \propto X_3^2$ is fully compatible with the KCM picture. Our explanation above show that this cannot be the case.

Let us repeat here Speck's argument for the sake of clarity: when $\omega \sim \tau^{-1}$, there is a number $\propto c(T)$ of active clusters of size $\ell(T)$ that can respond to an external electric field. Each such cluster carries a dipole $\sim \ell^{d/2}$ and thus, according to an argument very similar to the one leading to Eq. \eqref{eq2}, the $k$-th order peak susceptibility should scale as:
\begin{equation}\label{chi_kcm}
    \chi_k^{\text{\sc{kcm}}} \sim \frac{\mu^{k+1}}{T^k} c(T) \, [\ell(T)]^{d(k-1)/2},\qquad (k=1,3,5,\cdots)
\end{equation} 
from which, using 
\begin{equation} \label{eq_X_kcm}
   X_3^{\text{\sc{kcm}}}:=T \frac{\chi_3^{\text{\sc{kcm}}}}{(\chi_1^{\text{\sc{kcm}}})^2}; \qquad X_5^{\text{\sc{kcm}}}:=T^2 \frac{\chi_5^{\text{\sc{kcm}}}}{(\chi_1^{\text{\sc{kcm}}})^3}, 
\end{equation}
one obtains $X_3^{\text{\sc{kcm}}} \sim \ell^d/c$ and $X_5^{\text{\sc{kcm}}} \sim \ell^{2d}/c^2$, and hence $X_5^{\text{\sc{kcm}}}\approx  (X_3^{\text{\sc{kcm}}})^\gamma$ with $\gamma \equiv 2$, as found in experiments. 

There are various problems with this argument. First, as argued above, it is not because domains are dynamically correlated that response to an external field is collective. In theories and models based on Kinetic Constraints, rearrangements are synchronized in time, but still local in space: there is no ``super-dipole'' of size $\ell^{d/2}$ which responds rigidly. But more mundanely, Eq. (\ref{chi_kcm}), which is crucial to obtain $X_5^{\text{\sc{kcm}}} = (X_3^{\text{\sc{kcm}}})^2$, also predicts that the linear susceptibility $\chi_1$ is proportional to $c(T)/T$ and should thus strongly decay as temperature is decreased, at variance with empirical results where $\chi_1$ increases roughly as $\sim 1/T$. If one (rightly) argues that $\chi_1$ is in fact dominated by immobile regions and not by active regions,\footnote{Note in that respect that $\chi_1$ is indeed {\it age independent}, whereas $\chi_3$ does depend on age \cite{Bru12}.} then the definitions of $X_3$ and $X_5$ in Eq. \eqref{eq_X_kcm} are inadequate and should rather read
\begin{equation}
     X_3=T^3 \, {\chi_3^{\text{\sc{kcm}}}}; \qquad X_5=T^5 \, \chi_5^{\text{\sc{kcm}}},
\end{equation}
where we have used $\chi_1 \sim \mu/T$. Now, assuming that only  active regions contribute to $\chi_3$ and $\chi_5$ as in Eq. (\ref{chi_kcm}) one finds that $\ell$ factors cancel out and
\begin{equation} 
   \frac{X_5}{X_3^2} = \frac{1}{c(T)}, 
\end{equation}
which should be (i) quite large since the concentration of defects $c(T)$ is assumed to be small in the glassy region; and more importantly (ii) strongly increase as temperature is reduced
since $c(T)$ is assumed to decrease considerably when temperature is lowered. Both facts are in complete contradiction with our experimental data: $X_5/X_3^2$ is instead found to be roughly independent of temperature and at most equal to $\approx 2.5$.  

Hence, we cannot subscribe to the claim that kinetic pictures ``explain'' the experimental results on non-linear susceptibilities. The same remark applies to other non-thermodynamic theories of glasses, such as the ``shoving model''  \cite{shoving}. In fact, as we have argued elsewhere \cite{crumbling}, RFOT describes the very mechanism that leads to the appearance of local rigidity, and hence to the possibility of extremely long-lived metastable states in these systems for $T < T^\star$ or $T_o$. Absent such a local caging mechanism that hinders individual moves, the whole idea of kinetically constrained models falls apart. 

Therefore, theories based on Kinetic Constraints should first come up with a consistent scenario for local rigidity -- like for example the one proposed in \cite{CW} -- before invoking facilitation as the mechanism for sluggishness. Whereas facilitation processes are no doubt present in glasses, we believe that the fundamental mechanism leading to metastability for $T < T_o$ is of thermodynamic origin, and is necessarily accompanied by some incipient amorphous order. 

On the other hand, dynamical heterogeneities -- characterizing how many molecules rearrange in a {\it correlated} way, but not necessarily {\it cooperatively} -- do extend over length scales that are somewhat larger than the length $\ell$ over which amorphous order sets in. This is trivially true for KCM, since dynamical correlations grow in the absence of any local order. More generally, it was shown in \cite{Bou05} that the dynamical correlation length can only diverge faster than the amorphous order correlation length $\ell$ that governs non-linear susceptibilities.

Finally, we do not agree either with the argument given by Speck \cite{speck} on the role of random pinning on glassiness induced by Kinetic Constraints. Since the size of wandering defects is much smaller than $\ell$, pinning a fraction $c_{\mathrm{pin}} \sim \ell^{-d} \ll 1$ of particles should not affect the overall dynamics, contrary to what is assumed by Speck. In fact any wandering defect can easily by-pass the few pinned particles. Indeed, the numerical study of the plaquette models in which Kinetic Constraints affect the dynamics does not show any dramatic effect on the relaxation time of pinning \cite{jack}. Hence, in absence of a precise analysis or detailed model studies, we do not see how theories only based on Kinetic Constraints can explain the results of, e.g. Ref. \cite{Oza15}, where randomly pinning a small fraction of particles dramatically changes the glassy dynamics. 
In our view, random pinning is instead a promising avenue for gaining more insights about the glass transition,  the ideal glass phase and the validity of the RFOT theory. 

\section{Conclusion \& Open Issues} 

We have shown that a growing static length associated to an emerging amorphous order naturally leads  to growing non-linear susceptibilities. This generally holds true when there is an underlying growing static length, as the one envisioned in RFOT theory but also in the case of locally preferred structures \cite{coslovich} or frustration-limited domains \cite{Tar05}. In contrast, we have argued that a purely dynamical mechanism for the slowing-down of the dynamics, such as the one put forward in Kinetic Constraints based theories \cite{KCM}, cannot be responsible for the growth of the non-linear dynamical susceptibilities, mainly because there is no spatial propagation of information in such a framework. It would actually be quite instructive to study numerically the non-linear susceptibilities of model systems and confirm that KCMs do behave trivially in that respect, whereas more realistic models indeed exhibit an increase of non-linear response as the glass transition is approached.

The experimental results reviewed in this paper clearly demonstrate that non-linear susceptibilities do grow as the system slows down. The growth is stronger for higher non-linearities, as expected for a critical phenomenon. This is solid evidence that the glass transition is accompanied by a growing static length of the type predicted by RFOT theory. Such a conclusion however does {\it not} contradict the idea that dynamical facilitation plays a role in the dynamics of supercooled liquids. Even within RFOT, it is clear that the relaxation of a given glassite will trigger relaxation in neighbouring glassites, and hence that dynamical correlations must extend beyond the size of the glassites $\ell$ (see \cite{Xia}, and \cite{crumbling} for a recent discussion of this point). Whereas non-linear susceptibilities probe the increase of $\ell$, dynamical correlations should extend over a possibly much larger length-scale.  


Hence the question is not whether dynamical facilitation, rather than amorphous order, can explain the increase of the non-linear susceptibilities. As the theoretical and experimental results presented here show, such a growth must be due to an increase of a static length. The central issue is instead assessing the relative role of dynamical constraints versus amorphous order in the momentous increase of the relaxation time in glassy system. [Note that such an increase in time-scale amounts to a multiplication of the activation energy, for fragile glasses, by a factor $2$ to $5$ between the incipient rigidity temperature $T^\star$ and the glass transition temperature $T_g$].

This question has been debated recently in relation with the spectacular acceleration of ``swap'' Monte Carlo simulations \cite{swap}; we refer to \cite{CW,crumbling} for two opposite conclusions on this matter. In order to settle this issue, the most pressing theoretical is to elicit the precise nature and geometry of the activated events that unlock the amorphous order inside glassites \cite{Bou04,BCX}. In particular, a quantitative understanding of the activation energy, and its dependence on the size of glassites $\ell$ \cite{Cam09}, is still very much needed to patch together all the pieces of the puzzle -- in particular the power-law behaviour of $\chi_3$ and $\chi_5$ for $\omega \tau \gg 1$. Note that the aging experiments of \cite{Bru12} offer some important insights, again compatible with the prediction of RFOT.

Finally, let us end where we started, i.e. with spin-glasses. Although many experiments have been performed to probe the divergence of non-linear susceptibilities above the phase transition, and the aging properties of the linear susceptibility below the transition, we are not aware of any experimental investigation of the non-linear susceptibility in the aging regime. This would provide very interesting information on the nature of the spin-glass phase, and, possibly, an alternative approach to the vexing question of the existence of a de Almeida-Thouless line in three dimensional spin-glasses (on this point, see the discussion in \cite{Bou05}). 

\paragraph{Acknowledgments} We thank Dave Thirumalai for inviting us to put together our thoughts and write this piece. We also want to acknowledge the tremendous influence of his (and his collaborators') ideas on our own research trail. We also thank all our collaborators on these topics for their numerous and invaluable inputs.

\end{document}